\documentclass[aps,prl,twocolumn,groupedaddress,showpacs]{revtex4-1}
\usepackage{mathbbol,amsmath,amsfonts,amssymb,bm,color,dsfont}
\usepackage{graphicx,psfrag}

\def\d{{{\sf d}}}
\def\r{{\bm{r}}}

\def\i{{\bm{e_1}}}
\def\j{{\bm{e_2}}}

\def\tr{{{\sf Tr}}}
\def\trr{{{\sf tr}}}

\def\pf{\mathcal{Z}}

\providecommand{\ignore}[1]{}
\providecommand{\aucmnt}[1]{#1}
\renewcommand{\aucmnt}[1]{}

\begin{document}

\title{The non-Abelian Duality Problem}

\author{E. Cobanera}
\email[Electronic address: ]{ecobaner@indiana.edu}
\author{G. Ortiz} 
\affiliation{Department of Physics, Indiana University, Bloomington,
IN 47405, USA}
\author{E. Knill}
\affiliation{National Institute of Standards and Technology, Boulder, CO 80305, USA}

\date{\today}

\begin{abstract}
We exploit a new theory of duality transformations to construct dual
representations of models incompatible with traditional duality
transformations.    Hence we obtain a solution to the long-standing
problem of non-Abelian dualities that hinges on two key observations:
(i) from the point of view of dualities, whether the group of
symmetries of a model is or is not Abelian is unimportant, and (ii) the
new theory of dualities  that we exploit includes
traditional duality transformations, but also introduces in a natural
way more general transformations.
\end{abstract}
\maketitle

{\it Introduction.}--- Dualities have been recognized as powerful
non-perturbative mathematical tools to study strongly interacting
systems since Kramers and Wannier introduced them to determine the
exact critical temperature of the planar Ising model \cite{KW}.
\emph{Traditional dualities} (TD) as described in
Refs. \cite{savit,druhl_wagner,malyshev_petrova} are obtained by a
systematic method based on the Fourier transform (FT), suitably
generalized to arbitrary groups \(G\). The method generates a dual
partition function (or lattice Euclidean path integral)
\(\pf^D[K_i^*]\) from a partition function (PF) \(\pf[K_i]\) with
physical couplings \(K_i,\ i=1,\cdots,m\). The dual PF has the
remarkable property that its (dual) couplings \(K_i^*\) are large
(strong) if the couplings \(K_i\) are small (weak), and {\it vice
  versa}. This is in part because the duality engenders collective
(topological) excitations in terms of which \(\pf^D\) is expressed.

Unfortunately, many models of great physical interest such as
Heisenberg, non-Abelian gauge and more recent models based on Hopf
algebras are outside the scope of TD transformations. The reason is
technical, not physical: the group-theoretic FT has different
algebraic properties depending on \(G\) being Abelian or not, and the
TD transformation takes advantage of essential simplifications present
only in the Abelian case.  In essence, a TD transformation introduces,
via an FT, dual elementary degrees of freedom (EDFs).  For Abelian
FTs, the dual EDFs are still locally coupled and result in physical
dual PFs. Non-Abelian FTs result in non-local interactions and/or
constraints and complex Boltzman weights, as historically illustrated
by attempts to construct dual representations of non-Abelian gauge
theories \cite{mandelstam}.  Thus, in order to obtain TDs, it is
necessary that the model and associate groups satisfy restrictive
properties enabling the existence of physical dual models.  

Conventionally it is thought that the group \(G\) needed for TD
transformations is determined by the model's group of symmetries $\cal
G$~\cite{savit,alvarez} (see especially section 7, point (3) of
Ref.~\cite{alvarez}).  Here we argue that \(G\) is not determined by,
and in general is unrelated to, $\cal G$.  Rather, \(G\) is associated
with and constrained by the model's local or quasi-local interactions.
 We call a model \emph{{\rm
    S}-Abelian}, or \emph{{\rm S}-non-Abelian}, according to whether the
group of symmetries $\cal G$ is Abelian or not.  Many models are
S-non-Abelian, but have a TD transformation with an associated Abelian
\(G\).  It is tempting to call a candidate duality transformation
D-Abelian or D-non-Abelian according to whether \(G\) is Abelian or
not. However, the underlying group may not be apparent and may involve
more general structures.  Instead, we focus on the presence or absence
of non-trivial constraints on the states of the models. That is, we
say that a \emph{transformation} connecting two locally defined PFs
has \emph{{\rm D}-non-Abelian} features if the transformation
introduces or removes non-trivial local constraints.  From this
perspective, it is impossible to have a D-non-Abelian \emph{
  self-duality}.

{\it The non-Abelian duality problem} is the problem of extending the
scope of TDs without sacrificing their physical content to cases where
there are no relevant Abelian groups \(G\) for the interactions of a
model.  Our main contribution is to introduce a generalization of TD
transformations, \emph{bond-algebraic duality transformations}, that
addresses the problem of non-Abelian dualities by exploiting the
recently developed theory of bond algebras~\cite{ba, con} and their
homomorphisms. These transformations~\cite{conII} handle on equal
footing models with arbitrary \(G\), Abelian or not, and even more
general models, where there is no obvious group structure constraining
the transformations. Unlike a strictly D-Abelian duality, a
bond-algebraic duality can have both D-Abelian and D-non-Abelian
features.  To illustrate our ideas, we give a duality for a model
outside the scope of TDs, namely a rigid-rotator model with group
\(G=SU(2)\). According to our terminology, this duality is
D-non-Abelian and impossible to obtain by a TD.  

{\it Lattice Models.---} For simplicity, consider models with
identical, classical EDFs with configuration space \(M\) at sites
\(\r\) of a lattice $\Lambda$. A full configuration of the model
consists of an assignment \(s_\r\in M\) for each site $\r$.  If the
model has only pair-wise symmetric interactions, then the total energy
\(E\{s_\r\}\) of a configuration \(\{s_\r\}\) is a sum of (oriented)
two-body interaction energies \(\epsilon(s_\r,s_{\r'})=
\epsilon(s_{\r'},s_{\r})\).    This minimal description suffices
to specify physical quantities such as a PF. 
However, it often happens that \(M\) admits useful
additional mathematical structures.  In the context of TDs, this
includes groups acting on the EDFs.  
 More generally, we can consider configuration spaces that
are endowed with two operations \(m,m'\mapsto m \cdot m'\)
(multiplication) and \(m\mapsto S(M)\) (involution) such that (a)
multiplication is associative, (b) \(S\) is involutive (\(S^2\) is the
identity map) and order-reversing (\(S(m \cdot m')=S(m')\cdot S(m)\)),
and c) the pair-wise interactions between EDFs can be expressed in the
form
\begin{eqnarray}\label{compatiblemu}
\epsilon(s_\r,s_{\r'})=v\big(s_\r \cdot S(s_{\r'})\big),
\end{eqnarray}
for some real-valued function $v$.  Conditions (a) and (b) turn \(M\)
into a semigroup with involution.  We call models satisfying these
conditions \(m\)-models (short for {\it multiplication}-models).  It
is possible to accommodate interactions involving more than two EDFs,
provided the EDFs in an interaction are ordered and oriented. For
example, let \(s_{\r_1}, s_{\r_2},s_{\r_3},s_{\r_4}\) occupy the
corners of an elementary plaquette on the lattice, ordered along the
boundary of the plaquette. Then
\begin{eqnarray}\label{4bodym}
\epsilon(s_{\r_1},s_{\r_2},s_{\r_3},s_{\r_4})=v\big(s_{\r_1}\cdot S(s_{\r_2}) 
\cdot s_{\r_3}
\cdot S(s_{\r_4})\big)
\end{eqnarray} 
describes a form of \(m\)-interaction relevant to physical
applications that we discuss in the next section.

Wilson's lattice approach to quantum field theory \cite{wilson}
popularized the study of \(m\)-models defined in terms of EDFs taking
values on a group \(G=M\), with interactions of the form of
Eq. \eqref{compatiblemu} or its generalizations. These {\it
  \(G\)-models} are important examples of \(m\)-models where the
multiplication in $M$ is group multiplication and \(S\) is group
inversion, \(S(g)=g^{-1}\). \emph{ TD transformations are applicable
  only to \(G\)-models with \(G\) an Abelian group}
\cite{druhl_wagner}.  A reason for introducing the more general notion
of \(m\)-model is that we want to accommodate a larger set of theories,
such as those based on general Hopf algebras \cite{kitaev} that are
becoming increasingly more important in topological quantum matter,
and the theory of quantum computation and error correction.  

A model's symmetry group \({\cal G}\) is completely determined by its
interactions.  But semigroups with involution \(M\) associated with
the model and constrained to satisfy identities such as those of
Eqs. \eqref{compatiblemu} or \eqref{4bodym} are in general not unique
and may be completely unrelated to \({\cal G}\). For example, consider the
non-Abelian group \(S_N\) of permutations on \(N\geq 3\) letters, and
use it as the configuration space \(M=S_N\) for the EDFs of the Potts
model. Then we can write the interaction energy as
\begin{equation}\label{potts}
\epsilon_{\sf Potts}(s_\r,s_{\r'})=\delta_e\big(s_\r \cdot s_{\r'}^{-1}\big) ,
\end{equation}
where \(\delta_{e}(g) \! =\! \delta_{e,g}\) is the Kronecker delta on
\(S_N\).  The Potts model is non-Abelian from the point of view of its
symmetries, but it supports D-Abelian dualities.  
  The reason is that we can map the elements of \(S_N\)
to the elements of \(\mathbb{Z}_{N!}\) (the Abelian group of integers
modulo \(N!\)), and rewrite the interaction energy in the equivalent
form \(\epsilon_{\sf Potts}(s_\r,s_{\r'})=\delta_0\big(s_\r
-s_{\r'}\big)\).  Rewriting the model in this way does not change the
fact that its symmetries are non-Abelian, yet it permits the use of a
TD to determine its critical coupling. Some early explorations of
non-Abelian dualities \cite{zm, drouffe, orland} exploited this
procedure extensively to map models defined on certain non-Abelian
groups to Abelian ones.  
 In particular, it was noted that models defined on solvable
groups are specially amenable to this procedure \cite{drouffe}, since
solvable groups can be mapped to Abelian groups in a natural way.

{\it Beyond traditional dualities.}--- The recently developed theory
of bond-algebra homomorphisms \cite{con, conII} includes and
generalizes the theory of TD transformations.  To apply this theory,
we start with a physical model defined by its EDFs and local
interactions that capture the main features of the physical phenomena
under study. We then identify the model's \emph{bonds}, which are the
local or quasi-local interaction operators occuring in the
interactions. The multiplicatively closed algebra generated by the
bonds is called the \emph{bond algebra}. A key observation is that the
structure of the bond algebra and its generating bonds contain
essential information about the model. In particular, mappings between
bond algebras \emph{that preserve locality in, and all the algebraic
  relations among the bonds}, can demonstrate close relationships
between seemingly unrelated models, including models with EDFs of
differing exchange statistics. Although such bond-algebra mappings are
by definition local {\it in the bonds}, they are typically
\emph{non-local in the EDFs}. That is, the model's EDFs in the domain
can be naturally related in the range to highly non-local degrees of
freedom involving many EDFs~\cite{con, conII}.  These collective modes
can be considered to be alternative EDFs relative to which
interactions take different, but still local, forms.  In the
following, we call mappings of bond algebras that preserve locality
and algebraic relationships {\it bond-algebraic duality
  transformations}.  This is motivated by the observation made in
Refs. \cite{con, conII} that they can be used as the foundation for a
unified theory of classical and quantum dualities. Here we show that
bond-algebraic dualities go beyond TDs and generate new
transformations that are not related to the group-theoretic FT.

A bond-algebraic duality \cite{conII} for a classical model can be
obtained by expressing the PF \(\pf\) in terms of operators that can
be related to a bond algebra.  A popular way to do this (for an
alternative, see ~\cite{somma}) begins by identifying operators
\(T_0,\cdots, T_s\), called transfer matrices (TMs), acting on a
Hilbert space \(\mathcal{H}\), and a preferred basis
\(\phi=\{|\phi_i\rangle\}\) of \(\mathcal{H}\). The operators must
satisfy
\begin{equation}\label{1}
\pf=\sum_{\{s_\r\}} [e^{- E\{s_\r\}}]= \tr_{\phi}[(T_s \cdots T_1T_0)^N],
\end{equation}
where \(N\) is determined by the length of the lattice in a chosen
direction. The role of the basis is so enable us to make the equality
explicit by appropriately inserting resolutions of the identity
$\sum_i|\phi_i\rangle\langle\phi_i|$ between the operators in the
trace, expanding the trace in terms of the resulting summands and
associating the states $s_\r$ with sequences of basis indices. For
this to work and the expanded trace to match the desired PF, we need
the right combination of TMs and a preferred basis.

The locality of the classical model's interactions is usually
reflected in this construction. Thus, the Hilbert space $\mathcal{H}$
is defined by quantum EDFs on a lattice such that the TMs factor into
a product of quasi-local operators, $T_\alpha=\prod_{\Gamma}
t_{\alpha, \Gamma}$ ($\alpha=0,1,\cdots,s$), with \(\Gamma\) a lattice
index that may stand for a site, a link, or a plaquette.  As a result, it
is natural to define the bond algebra of \(\pf\) as the algebra
generated by the {\it bonds} \(\{t_{\alpha, \Gamma}\}\)~\cite{ba}.

To obtain a duality, one can algebraically represent the bonds, and
therefore the TMs, on an alternative space, and determine a preferred
basis so that the expansion of the trace can be recognized as a
physical PF for a different model.  Suppose we have such a
bond-algebraic duality \(t_{\alpha, \Gamma}\mapsto t_{\alpha,
  \Gamma}^D\) with image bonds \(t_{\alpha, \Gamma}^D\) on different
quantum EDFs that are also local and have the same algebraic
relationships.  This induces a bond-algebra isomorphism between the
algebras generated by the two sets of bonds.  We can define dual TMs
\(T_\alpha^D=A_\alpha^{-1/N}\prod_\Gamma t_{\alpha, \Gamma}^D\), with
$A_\alpha$ analytic functions of the parameters of the model, and
compute a dual PF as
\begin{equation}\label{classicalbad}
\pf^D=\tr_\psi[(T_s^D \cdots T_1^D T_0^D)^N].
\end{equation}
relative to a basis \(\psi=\{|\psi_j\rangle\}\) to be specified.  A
nontrivial property of typical bond algebra isomorphisms is that they
are induced by unitary transformations~\cite{conII}.  In particular,
if \(t_{\alpha, \Gamma}^D =\mathcal{U}_\d t_{\alpha,
  \Gamma}\mathcal{U}_\d^\dagger\), with $\mathcal{U}_\d$ unitary, then
\begin{equation}\label{classduality}
\pf=A \ \pf^D\  \ \ \mbox{ and  }  \ A=\prod_{\alpha=0}^s A_\alpha .
\end{equation} 
It follows that \(\pf\) and \(\pf^D\) represent two, in general
different, systems that have nonetheless the same thermodynamics.

\begin{figure}[thb]
\centering
\includegraphics[angle=0,width=.7\columnwidth]{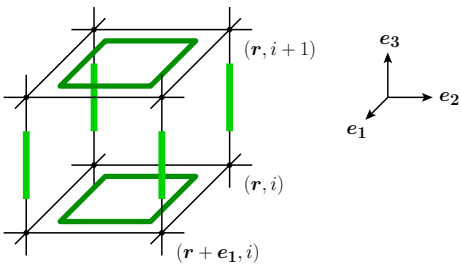}
\caption{Lattice connectivity of  the classical $D=3$ XM model.}
\label{classicalxm}
\end{figure}

The final form of \(\pf^D\) in terms of its EDFs depends critically on
the choice of basis \(\psi\) in Eq. \eqref{classicalbad}.  As an
extreme example, if \(\{|\psi_j\rangle\}=
\{\mathcal{U}_\d|\phi_i\rangle\}\), then Eq.~\eqref{classduality} is
reduced to a trivial identity with $A=1$. The choice of basis also
determines whether a bond-algebraic duality is D-non-Abelian or
D-Abelian, that is, whether or not it introduces local constraints
when the trace is expanded. Local constraints appear if the
combination of TMs between resolutions of the identity have entries
that are zero with respect to the basis. Thus, given a bond-algebraic
duality, it is natural to seek a basis where the relevant TMs are
full, so that the duality is D-Abelian.  In general, the entries of
the matrices also need to be positive and expressible as products of
local Boltzmann weights.  Although such bases are known to exist for a
large class of duality problems including TDs, we do not have general
strategies for finding them.

\begin{figure}[h]
\includegraphics[angle=0,width=.85\columnwidth]{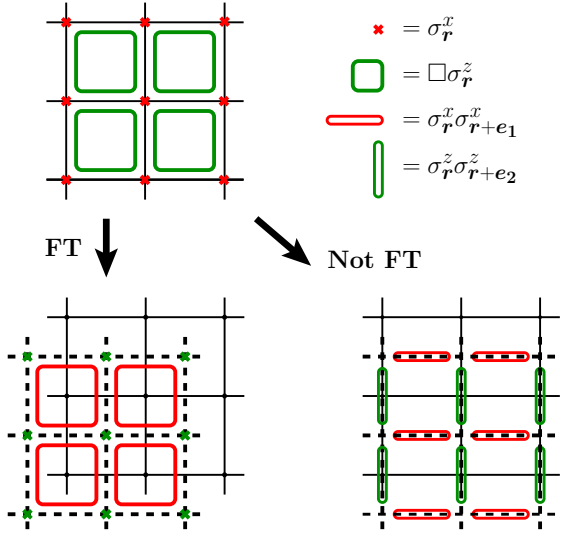}
\caption{The quantum XM model (shown on top) is self-dual as indicated
  by the arrow on the left, and it is dual to the planar orbital
  compass model, as indicated by the arrow on the right. Direct and
  dual lattices are indicated with solid and dashed lines,
  respectively.    }
\label{xmpoc}
\end{figure}

We illustrate these ideas with a D-Abelian and a D-non-Abelian duality
for the Xu-Moore (XM) model of \(p+ip\) superconducting arrays
\cite{xm_prl, xm_npb}. The model's \(D=3\) dimensional classical PF is
given by (see Fig. \ref{classicalxm})
\begin{equation}\label{classicalxmf}
\pf_{\sf XM}=\sum_{\{\sigma_{(\r,i)}\}}e^{\sum_i\sum_\r[K_l\sigma_{(\r,i)}
\sigma_{(\r,i+1)}+K_p\square\sigma_{(\r,i)}]}, 
\end{equation}
where \(\sigma_{(\r,i)}=\pm 1\) are classical Ising variables placed
at the sites \((\r,i)\) (\(i\) an integer) of a cubic lattice, and
\(\square\sigma_{(\r,i)}\equiv \sigma_{(\r,i)}\sigma_{(\r+\j,i)}
\sigma_{(\r+\i,i)}\sigma_{(\r+\i+\j,i)}\). The XM model is a
\(G\)-model with \(G=\mathds{Z}_2\). 
  The TD transformation maps the model to
itself with a characteristic interchange of strong and weak coupling
constants \cite{xm_prl}.  To recast it as a D-Abelian bond-algebraic
duality, we construct plane-to-plane TMs
\begin{equation}\label{txm}
T_1=\prod_\r e^{h \sigma^x_\r},\ \ \ \
T_0=\prod_\r e^{K_p\square\sigma^z_\r}, 
\end{equation}
with \(\sigma^{x,z}_\r\) Pauli matrices acting on quantum spins at
sites \(\r\) of a (\(d=2\)) square lattice, \(h=-\ln\tanh(K_l)/2\),
and \(\square\sigma^z_\r=
\sigma^z_\r\sigma^z_{\r+\j}\sigma^z_{\r+\i}\sigma^z_{\r+\i+\j}\) (see
Fig. \ref{xmpoc}).  To recover \(\pf_{\sf XM}\), the trace
\(\tr_\phi[(T_1T_0)^N]\) is computed with respect to the basis $\phi$
that diagonalizes the \(\sigma^z_\r\).

The TMs can be expressed as products of \(t_{1,\r}=
\cosh(h)+\sinh(h)\sigma^x_\r, t_{0,\r}=
\cosh(K_p)+\sinh(K_p)\square\sigma^z_\r\).  We therefore let the bonds
be \(\{\sigma^x_\r, \square\sigma^z_\r\}\).  They satisfy a bond-algebraic
duality induced by
\begin{equation}\label{sd_XM}
\sigma^x_\r\mapsto\square\sigma^x_\r,\ \ \ \ \square\sigma^z_\r\mapsto\sigma^z_{\r+\i+\j},
\end{equation} 
and illustrated in Fig. \ref{xmpoc}.
The dual TMs
\begin{equation}
T_1^D=\prod_\r e^{h \square\sigma^x_\r},\ \ \ \ 
T_0^D=\prod_\r e^{K_p\sigma^z_\r}, 
\end{equation}
are related to \(T_1,T_0\) by a unitary mapping.  If we expand
\(\pf^D_{\sf XM}=\tr_\phi[(T_1^DT_0^D)^N]\) with respect to \(\phi\),
we find that \(\pf^D_{\sf XM}\) contains local constraints, so that
the mapping of Eq. \eqref{sd_XM} is D-non-Abelian relative to
$\phi$. It is, however, \(D\)-Abelian in the basis \(\psi\) that
diagonalizes the \(\sigma^x_\r\), with respect to which we recover the
traditional self-duality of the XM model \cite{xm_prl, con}.  In the
bond-algebraic approach to dualities, the role of the FT is encoded in
the change of basis \(\phi\mapsto \psi\) realized by a direct product
of Hadamard operators $H$ satisfying \(H\sigma^zH=\sigma^x\).

The bond algebra of the XM model has another local representation
\cite{nf, con, conII},
\begin{equation}\label{xmtopoc}
\sigma^x_\r\mapsto\sigma^z_\r \sigma^z_{\r+\j},\ \ \ \ 
\square\sigma^z_\r\mapsto \sigma^x_{\r+\j} \sigma^x_{\r+\i+\j}.
\end{equation} 
The corresponding dual TMs
\begin{equation}
{\tilde{T}_1^D}=\prod_\r e^{h \sigma^z_\r\sigma^z_{\r+\j} },\ \ \ \ 
{\tilde{T}_0^D}=\prod_\r e^{K_p\sigma^x_\r\sigma^x_{\r+\i}},
\end{equation}
yield an alternative dual partition function \({\tilde{\pf}^D_{\sf
    XM}}=\tr_\phi[({\tilde{T}_1^D}{\tilde{T}_0^D})^N]\).  
With $\phi$ the basis that diagonalizes the \(\sigma^z_\r\), we obtain
a PF with {\it local, four-spin constraints}. Relative to this basis
the duality of Eq. \eqref{xmtopoc} is D-non-Abelian.  It is an open
problem whether there is a choice of basis for which
\({\tilde{\pf}^D_{\sf XM}}\) is an unconstrained canonical ensemble
making the duality D-Abelian.  An alternative
  may be to remove these constraints by reinterpreting them as gauge
  symmetries.

It is important to recall at this point that a TD maps a \(G\)-model
on a lattice \(\Lambda\) to an essentially unique dual model supported
on the dual lattice \(\Lambda^*\)~\cite{druhl_wagner}, and the XM
model is self-dual under such TDs.  In contrast, the bond-algebraic
duality of Eq.~\eqref{xmtopoc} results in a model with a Hamiltonian
that differs from that of the XM model.  We conclude that this
bond-algebraic duality is not a TD.

{\it Non-Abelian dualities.---} Next, we show that bond-algebraic
dualities exist for $G$-models with non-Abelian $G$ and no TDs.
 For example, consider the Euclidean
lattice version \cite{wilson} of the \(SU(2)\) principal chiral field
\cite{sigma}. This model involves an \(SU(2)\)-valued field
\(u(x)=
\begin{pmatrix}  u^1_{\ \, 1} & u^1_{\ \, 2} \\ u^2_{\ \, 1}  &u^2_{\ \, 2}  \end{pmatrix}
\in SU(2)\) with action
\begin{equation}
 S_{\sf PCh}=\frac{1}{2\lambda_0}
\int dtdx\ \trr(\partial_0u^*.\partial_0u-\partial_1u^*.\partial_1u)\ .
\end{equation}
The lower dot denotes matrix multiplication, \(u^*(x)\) is the
Hermitian-conjugate field, and \(\trr\) is the $2\times 2$-matrix
trace.  Since \(u^*(x)u(x)=\mathds{1}\), the lattice Euclidean path
integral reduces to
\begin{equation}\label{latticenlsm}
\pf_{\sf PCh}=\int_{\{u_\r\}}e^{\frac{1}{2\lambda_0}\sum_{\r}
{\sf Re}\{\trr(u^*_{\r+\i}u_\r)+\trr(u^*_{\r+\j}u_\r)\}}
\end{equation}
on the square lattice with $SU(2)$ as the EDFs' configuration
space. Note that if we replace \(SU(2)\) by \(U(1)\) we
obtain the \(XY\)-model, for which there is a $D$-Abelian duality to
the solid-on-solid model~\cite{savit, conII}.

To express \(\pf_{\sf PCh}\) in terms of row-to-row transfer
operators.  we use covariant pairs of standard representations of
\(SU(2)\) and the continuous functions \(C_0(SU(2))\) on \(SU(2)\),
both acting on wavefunctions on \(SU(2)\).  A generating set for
\(C_0(SU(2))\) is given by \((U)^\mu_{\ \, \nu}\) ($\mu,\nu=1,2$),
where $(U)^\mu_{\ \, \nu}(u) = u^{\mu}_{\ \, \nu}$. Thus $U$ is a
matrix-valued function. The standard representation of \(SU(2)\) has
infinitesimal generators $J=(J_x,J_y,J_z)$ for multiplication on the
right. If we write \(u=e^{-{\sf i}\theta
  \hat{n}\cdot\vec{\sigma}/2}\), $\theta$ a finite angle, and
$\hat{n}$ a unit vector, then $e^{{\sf i}\theta \hat{n}\cdot
  J}|v\rangle=|v.u\rangle $ for the formal basis of wavefunctions
$|v\rangle$.  The row-to-row transfer operators are given by
\begin{eqnarray}\label{ti}
T_0&=&\prod_{i}e^{\frac{1}{2\lambda_0} {\sf
    Re}\{\trr(U^*_i.U_{i+1})\}},\\ T_1&=&\prod_{i}\int du\ e^{h{\sf
    Re}\{\trr (u)\}} e^{{\sf i}\theta\hat{n}\cdot J_i}, \ \ u=e^{-{\sf
    i}\theta \hat{n}\cdot\vec{\sigma}/2},
\label{u}
\end{eqnarray}
for a parameter \(h\) dependent on \(\lambda_0\). The products are
over the EDFs in a row.  To recover Eq. \eqref{latticenlsm}, the trace
\(\tr_\phi[(T_1T_0)^N]\) is expanded with respect to the basis
$|v\rangle$. 

To define a bond-algebraic duality, we use the generators \(j_{i}\) of
left multiplication, which satisfy $e^{{\sf i}\theta \hat{n}\cdot
  j_i}|u_i\rangle=|u.u_i\rangle.$ These generators can be related to
actions defined by $J$ and $U$ by the identity \cite{kogut_susskind}
\( j_{ia}\equiv \sum_{b=x,y,z}\trr(U_i^*\sigma^aU_i\sigma^b) J_{ib}/2
\), such that \([j_i,J_j]=0\).  The bond algebra generated by the
local bonds \(J_i\) and \(U_i^*.U_{i+1}\) can be transformed to local
bonds according to
\begin{equation}\label{naigd}
J_{i}\mapsto -j_i+J_{i-1} ,  \ \ \ U_i^*.U_{i+1}\mapsto U_i .
\end{equation}
Proving that the mapping is induced by a unitary operator requires
adding boundary terms to complete the algebra, checking that the
images of the EDFs' operators are generated by corresponding
covariant pairs of representations and applying the Stone-von
Neumann-Mackey theorem~\cite{rosenberg_j:qc2004a} (see the
Supplemental Material).  It follows that
\begin{eqnarray}
T_0^{D}&=&\prod_i \ e^{\lambda_1{\sf Re}\{ \trr(U_i)
  \}}\ ,\\ T_1^D&=&\prod_{i} \int du \ e^{\lambda_2{\sf Re}\{
  \trr(u)\}} e^{{\sf i}\theta\hat{n}\cdot(-j_i+ J_{i-1})}
\end{eqnarray}
are unitarily equivalent to the corresponding TMs $T_0$ 
and $T_1$. Note that the dual variables \(\hat{J}_i, \hat{U}_i\)
\begin{eqnarray}\label{nadvars}
\hat{J}_i=-j_i +J_{i-1},\ \ \ \ \hat{U}_i= \cdots
.U_{i+2}^*.U^*_{i+1}.U_i^*
\end{eqnarray}
that are the unitary images of the EDF operators under the duality
are, as expected on general grounds, non-local collective modes. The
string defining \(\hat{U}_i\) extends to the boundary of the system,
and its specific form is determined by the chosen boundary conditions.

To obtain a dual PF, we expand the trace with respect to the basis
$|v_i\rangle$ for each $i$. The PF \(\pf_{\sf
  PCh}^{D}=\tr_\psi[(T_1^DT_0^{D})^N]\) is then given by
(\(\r=i\i+j\j\))
\begin{equation}\label{gdnai}
\int_{\{u_\r\}}
e^{\sum_{i,j}\frac{1}{2\lambda_0} {\sf Re}\{\trr(u_{i,j}^*.u_{i,j+1})+
\trr(u_{i,2j})\}} \!\! \prod_{i+j={\sf even}} \!\!\!
\delta(\mathds{1},\square u_{i,j}), \nonumber
\end{equation}
where $\square u_{i,j}=u_{i,j}^*u_{i,j+1}u_{i+1,j+1}u_{i+1,j}^*$ (see
the Supplemental Material).  As for \(\pf^D_{\sf XM}\), we obtain a PF
with {\it local constraints} on a checkerboard. We do not know
whether there is a choice of basis that removes such constraints.

We have discussed dualities for classical models, which also apply to
Euclidean path-integral representations of quantum problems.  This is
the context in which the problem of non-Abelian dualities is typically
stated.  However, as explained in detail in Ref. \cite{conII},
bond-algebraic dualities provide a unified approach to classical and
quantum dualities, so we can use essentially the same techniques
to obtain dualities for any quantum mechanical model. The bond algebra
of a quantum Hamiltonian \(H=\sum_\Gamma h_\Gamma\) is the algebra
generated by the local or quasi-local bonds \(h_\Gamma\), and a
bond-algebraic quantum duality is given by a mapping \(h_\Gamma\mapsto
h_\Gamma^D\) to an algebraically equivalent dual set of local or
quasi-local bonds. As before, one can typically show that the
isomorphism is induced by a unitary transformation, in which case
\(H^D=\sum_\Gamma h_\Gamma^D\) is unitarily equivalent to \(H\). Take,
for example, the $d=1$, infinite chain, $SU(2)$ equivalent of the $\mathds{Z}_2$
transverse-field Ising Hamiltonian
\begin{equation}
H_{\sf PCh}= \sum_i [\frac{1}{2}J_i^2+\frac{\lambda}{2} (\trr
  (U_i^*.U_{i+1})+\trr(U_{i+1}^*.U_i))] ,
\end{equation}
which is not self-dual, but has a duality to
\begin{equation}
H_{\sf PCh}^D=\sum_i [\frac{1}{2}(-j_i+J_{i-1})^2+\frac{\lambda}{2} 
(\trr (U_i^*)+\trr(U_i))] ,
\end{equation}
as follows from Eq. \eqref{naigd} (see the Supplemental Material). 
Quantum dualities are remarkably
simpler than classical dualities. They do not depend on a choice of
basis, and so the distinction between D-Abelian and D-non-Abelian
becomes irrelevant.

\acknowledgments
Contributions to this work by NIST, an agency of the US government,
are not subject to copyright laws.

\newpage
\section{Supplemental Material to {\it The non-Abelian Duality Problem}}

We present mathematical details and clarify technical issues of results reported in the
accompanying paper {\it The non-Abelian Duality Problem}.

\section{The Ising Model Revisited}

In this section we explain how to choose the basis for
expanding the traces when determining partition functions from
products of transfer matrices.  We illustrate the main concepts 
by example and use the two-dimensional Ising model~\cite{bookEPTCP} 
to show that bond-algebraic dualities may display both D-Abelian and 
D-non-Abelian features.  The
partition function of the Ising model is given by ($\r= i \i + j \j$)
\begin{equation}
\pf_{\sf I}[K_1,K_2]=\sum_{\{\sigma_{\r}\}} \exp[\sum_{i,j}(K_1\sigma_{i,j}\sigma_{i+1,j}
+K_2\sigma_{i,j}\sigma_{i,j+1})], 
\end{equation}
and can be expressed as 
\(
\pf_{\sf I}[K_1,K_2]=\tr_\phi [(T_1T_0)^N]
\)
in terms of the transfer matrices
\begin{equation}
T_0=\prod_i\ e^{K_1\sigma^z_i\sigma^z_{i+1}},\ \ \ \ \ \
T_1=\prod_i\ (e^{K_2}+e^{-K_2}\sigma^x_i),
\end{equation}
provided the trace is expanded in the basis
\(\phi=\{|\phi_k\rangle\}\) that diagonalizes the Pauli matrices
\(\sigma^z_i\). 

\begin{figure}[h]
\centering
\includegraphics[angle=0, width=.8\columnwidth]{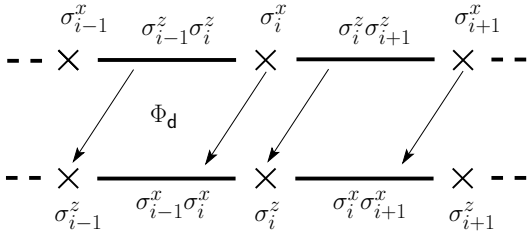}
\caption{Duality isomorphism of bond algebras associated with the
  transfer matrices of the Ising model.  }
\label{iid}
\end{figure}

The mapping 
\begin{equation}\label{isingdm}
\sigma^z_i\sigma^z_{i+1}\mapsto\sigma^z_i,\ \ \ \ \ \ 
\sigma^x_i\mapsto \sigma^x_{i-1}\sigma^x_i,
\end{equation}
illustrated in Fig. \ref{iid} defines an isomorphism of bond algebras
that is induced by a unitary mapping.  Thus \(T_0, T_1\) are dual and unitarily
equivalent to
\begin{equation}\label{iidtm}
T_0^D=\prod_i\ e^{K_1\sigma^z_i },\ \ \ \ \ \
T_1^D=\prod_i\ (e^{K_2}+e^{-K_2}\sigma^x_i\sigma^x_{i+1}).
\end{equation}

\begin{figure}
\centering
\includegraphics[angle=0, width=.6\columnwidth]{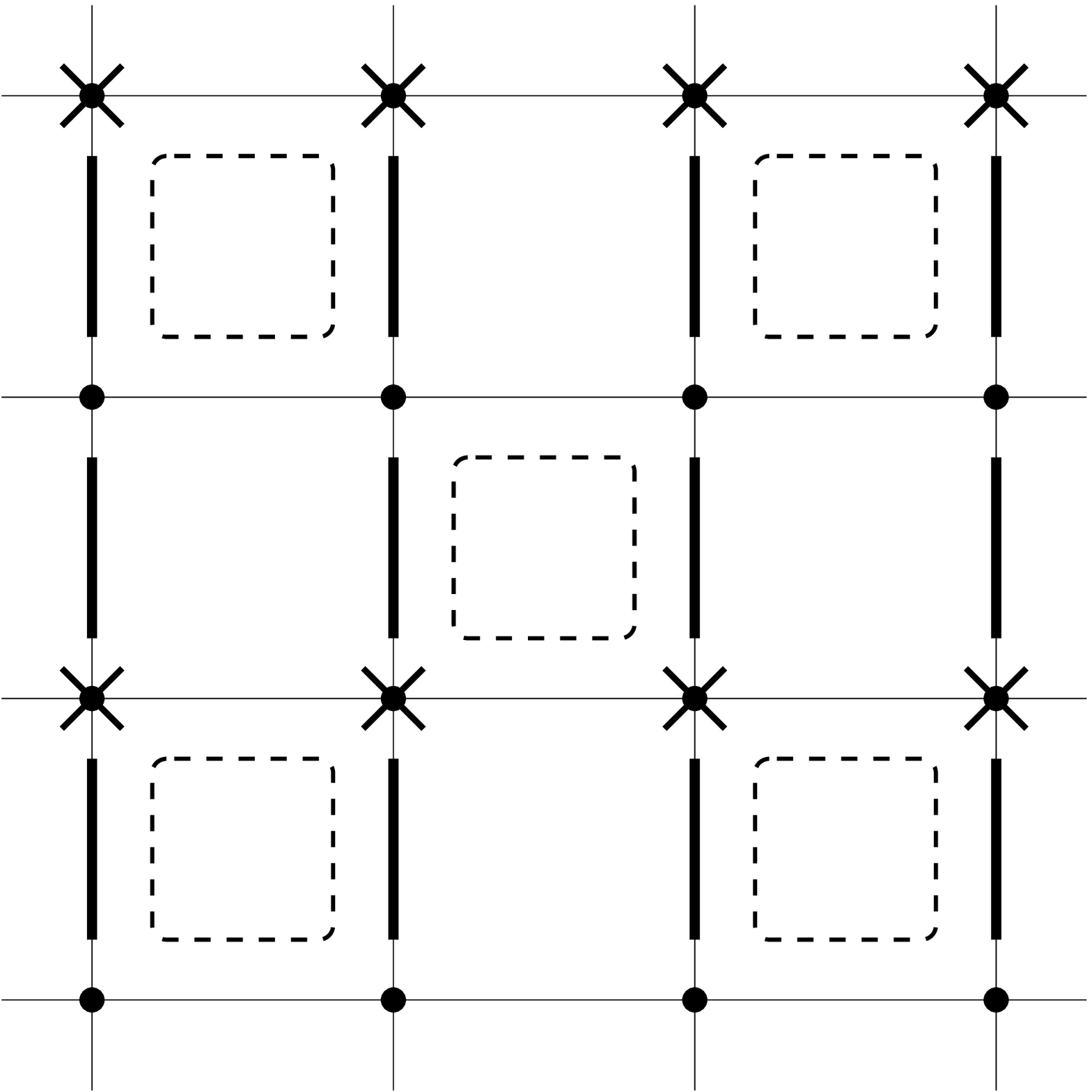}
\caption{Interactions and constraints in the dual partition function \({\pf}_{\sf I}^D\).
The crosses highlight the sites where the classical Ising variables couple to a inhomogeneous
external field of magnitude \(K_1\). The heavy vertical lines indicates a nearest-neighbor
Ising interaction of magnitude \(K_2/2\). The staggered distribution of plaquettes with round
corners indicates the distribution of four-spin delta constraints.
}
\label{ciid}
\end{figure}

To compute a partition function from the dual transfer matrices via
the expression \(\pf_{\sf I}^D=\tr_\psi [(T_1^DT_0^D)^N\)], we need to
specify a basis \(\psi=\{|\psi_k\rangle\}\).  The expansion of the trace
obtained by inserting resolutions of the identity with respect to this
basis must be recognizable as the partition function of a local
system. In particular, the coefficients of the expansion must be
non-negative, so that they can be written as Boltzman weights, and they
must be products of local terms consistent with the expansion.  For
example, set \(\psi=\phi\), the basis of the previous paragraph.
Then, as will become clear below, it is convenient to split
\({T}_1^D={T}_{1o}^{D}{T}_{1e}^{D}\), where
\begin{eqnarray}\label{iintd}
{T}_{1o}^{D}&=&\prod_{i_o}(e^{K_2}+e^{-K_2}\sigma^x_{i_o}\sigma^x_{i_o+1}), \\
{T}_{1e}^{D}&=&\prod_{i_e}(e^{K_2}+e^{-K_2}\sigma^x_{i_e}\sigma^x_{i_e+1}),
\end{eqnarray}
with $i_o=2i+1$, $i_e= 2i$, $i\in\mathds{Z}$.
We can label the members of the basis $|\phi_k\rangle$ in terms
of strings $\sigma$ of Ising variables $\pm 1$ at sites $i$
so that 
\(
\sigma^z_i|\sigma\rangle=\sigma_{i}|\sigma\rangle\ .\)
With these
labels, the basis members are written as $|\sigma\rangle$.
We now compute
\(
\pf_{\sf I}^D=\tr_\phi [({T}_{1o}^{D}{T}_{1e}^{D}T_0^D)^N]
\)
by expanding the trace as 
\begin{eqnarray}\label{ittt}
\tr_\phi [({T}_{1o}^{D}{T}_{1e}^{D}T_0)^N]&=&\\
 & &\!\!\!\!\!\!\!\!\!\!\!\!\!\!\!\!\!\!\!\!\!\!\!\!\! \sum_{\{\sigma_1\},\cdots,\{\sigma_{2N}\}} \!\!\!\!\!\!
\langle \sigma_{1}|{T}_{1o}^{D}|\sigma_2\rangle\langle \sigma_2|
{T}_{1e}^{D}T_0^D|\sigma_3\rangle
\cdots \nonumber \\
& \cdots& \langle \sigma_{2N-1}|{T}_{1o}^{D}|\sigma_{2N}\rangle
\langle \sigma_{2N}|{T}_{1e}^{D}T_0^D|\sigma_{1}\rangle \nonumber,
\end{eqnarray}
where \(\{\sigma_j\}, j=1,\cdots,2N\), describes the state of row
\(j\).  Note that \(T_0^D\) is diagonal in the chosen basis.  Further,
\begin{eqnarray}
\lefteqn{\langle \sigma_{j}|{T}_{1o}^{D}|\sigma_{j+1}\rangle=} \\
&\ & \prod_{i_o}\langle \sigma_{i_o,j}\sigma_{i_o+1,j}|
e^{K_2}+e^{-K_2}\sigma^x_{i_o}\sigma^x_{i_o+1} |\sigma_{i_o,j+1}\sigma_{i_o+1,j+1}
\rangle, \nonumber
\end{eqnarray}
and a similar factorization holds for \(\langle
\sigma_{j}|{T}_{1e}^{D}|\sigma_{j+1}\rangle\).  The splitting
\({T}_1^D={T}_{1o}^{D}{T}_{1e}^{D}\) was introduced to ensure this
factorization. We can evaluate Eq. \eqref{ittt} by applying shifted
forms of the identity
\begin{eqnarray}
\lefteqn{\langle \sigma_1'\sigma_2'|e^{K_2}+e^{-K_2}\sigma^x_{1}\sigma^x_{2}
|\sigma_1 \sigma_2\rangle=} \nonumber \\
&=&e^{\frac{K_2}{2}(\sigma_1'\sigma_1+\sigma_2'\sigma_2)}
\delta(\sigma_1'\sigma_1,\sigma_2'\sigma_2) . 
\end{eqnarray}
It follows that 
\begin{eqnarray}
{\pf}_{\sf I}^D&=&\sum_{\{\sigma_{\r}\}}
\left(\prod_{i+j={\sf even}} \delta(\sigma_{i,j}\sigma_{i,j+1},\sigma_{i+1,j}
\sigma_{i+1,j+1})\right)\nonumber\\
&&\times\exp \Big[\sum_{i,j}\big(\frac{K_2}{2} 
\sigma_{i,j}\sigma_{i,j+1}+K_1\sigma_{i,2j+1}\big) \Big].\label{ci}
\end{eqnarray}
The interactions in \({\pf}_{\sf I}^D\) are illustrated in Fig. \ref{ciid}.

The last factor in Eq. \eqref{ci} for \({\pf}_{\sf I}^D\) can be
identified as a Boltzmann weight for a physical system with local
interactions, one of the requirements for a good choice of basis to
expand the trace in.  However, the expression for the partition
function in Eq.~\eqref{ci} also introduces local (delta function)
constraints to account for the fact that the dual Boltzmann weights
vanish for some configurations.  It is preferable to find a basis
\(\psi\) where all the Boltzmann weights are strictly positive so that
there are no constraints. For the Ising model, one can find such a
basis by inspection.  Let \(\psi\) be the basis that diagonalizes the
Pauli matrices \(\sigma^x_i\). Then one can check that
\begin{eqnarray}
\pf_{\sf I}[K_1,K_2]&=&\tr_\phi[(T_1T_0)^N] \nonumber \\
&=&\tr_\psi[(T_1^DT_0^D)^N]=A \pf_{\sf I}[K_1^*,K_2^*]
\end{eqnarray}
with \(\sinh(2K_1^*)\sinh(2K_2)=1=\sinh(2K_2^*)\sinh(2K_1)\). 
The proportionality factor \(A\) is an analytic function
of the couplings and size of the system \cite{conII}.  We then
recover the Kramers-Wannier self-duality of the Ising model.

The duality of the Ising model
expressed by Eq. \eqref{ci} is {\it not} a self-duality. The
Kramers-Wannier self-duality as derived above is the result of
combining the bond-algebraic mapping of Eq. \eqref{isingdm} with a
suitable choice of basis \(\psi\). The dual
partition function according to Eq.~\eqref{ci} has restructured the
interactions drastically, but has left the couplings \(K_1, K_2\)
essentially unchanged.  Nevertheless, such dualities reveal key
properties of traditional dualities.   For
example, consider the two-point correlator \(\langle
\sigma_{m',n'}\ \sigma_{m,n}\rangle\). In the limit in which
\((m',n')\) is infinitely far from \((m,n)\), this correlator defines the 
square of the order parameter.
We can compute the correlator in the dual model of Eq.~\eqref{ci} as
\begin{eqnarray}
&&\langle \sigma_{m',n'}\ \sigma_{m,n}\rangle= \label{tpfdwlls}\\
&=&\frac{\tr_\phi [ T^{(N-n')}\ \sigma^z_{m'}\ T^{(n'-n)}\ \sigma^z_m\ T^n]}{\tr_\phi [T^N]}
\nonumber\\
&=&\frac{\tr_\phi [(T^D)^{(N-n')}\ \mu^z_{m'}\ (T^D)^{(n'-n)}\ 
\mu^z_m\ (T^D)^{n}]}{\tr_\phi [(T^D)^{N}]}
\nonumber\\
&=& \langle \mu_{m',n'}\ \mu_{m,n}\rangle, \nonumber
\end{eqnarray}
where \(T=T_1T_0\), \(\ {T}^D={T}_1^D{T}^D_0\) (see
Eq. \eqref{iintd}), and \(\sigma^z\mapsto \mu^z_m=
\sigma^z_m\sigma^z_{m+1} \sigma^z_{m+2}\cdots,\) the dual
Pauli spin operator from Eq.~\eqref{isingdm}. Hence 
\begin{equation}\label{defmu}
\mu_{m,n}=\sigma_{m,n}\sigma_{m+1,n} \sigma_{m+2,n}\cdots\ .
\end{equation}
In the dual model \({\pf}_{\sf I}^D\), the string 
correlator \(\langle \mu_{m',n'}\ \mu_{m,n}\rangle\) is (in the limit
of infinite separation) the square of the order parameter.
Thus, for example, if \(K<K_c\), 
\(\pf_{\sf I}\) is in its ferromagnetic phase, corresponding by duality
to a phase of \({\pf}_{\sf I}^D\) dominated by strong
correlations of string collective modes.

\section{Duality of the \(SU(2)\) Principal Chiral field: Hamiltonian Formulation}

This section discusses a duality for the finite system
\begin{equation}\label{hpch}
H_{\sf PCh}=\frac{1}{2}\sum_{m=1}^N J_m^2 +\frac{\lambda}{2} \sum_{m=1}^{N-1}
{\sf Re}\,\trr(U_{m+1}^*.U_m).
\end{equation}
The Hamiltonian \(H_{\sf PCh}\) can be obtained as the 
{\it time-continuum limit} \cite{fradkin_susskind,kogut_revI}
of the partition function of Eq. (\(14\)) of the accompanying paper.

\subsection{Algebra of a Single Quantum Rigid Rotator}

The kinematical algebra of a
rigid rotator \cite{kogut_susskind1} 
is defined by the relations among the canonical variables 
\(J_a, a=x,y,z\), \(U^\mu_{\ \ \nu}, \mu,\nu=1,2\),
\begin{eqnarray}
{}J_a^\dagger&=&J_a, \label{algJad} \\
{}[J_a,J_b]&=&i\epsilon_{abc}J_c, \label{algJcom}\\
{}[J,U]&=&\frac{1}{2}U.\sigma, \label{algJU}\\  
{}U^*.U&=&U.U^*=\mathds{1}, \label{algU}
\end{eqnarray}
introduced in the accompanying paper. Here
$\sigma$ denotes a standard Pauli matrix. 
The low dot denotes matrix multiplication to distinguish it
from tensor multiplication, and a centered dot denotes the 
standard Euclidean inner product. For example, \(J\cdot J=J_x^2+J_y^2+J_z^2\), and
\begin{equation}
[J,U]=\frac{1}{2}U.\sigma\ \leftrightarrow\ [J_a, U^\mu_{\ \ \nu}]=
\frac{1}{2}\sum_{\kappa} U^\mu_{\ \ \kappa}\sigma^\kappa_{a\ \nu}\ .
\end{equation}
Eqs. \eqref{algJad} and \eqref{algU} imply
\(
{}[J,U^*]=-\sigma.U^*/2.
\)

The algebra above affords a set of position-like operators \(U, U^*\)
and conjugate momenta \(J_a\) that suffice to specify completely the kinematics
of quantum tops. It is useful however to introduce three additional operators
\begin{equation}\label{jfromJU}
j_a \equiv \frac{1}{2}\sum_{b} \trr(U^*.\sigma_a.U.\sigma_b)J_b,\qquad a=x,y,z
\end{equation}
or just \(j=\trr(U^*.\sigma.U.(\sigma\cdot J))/2\) for short,
having some very useful properties: 
\begin{eqnarray}
{}j_a^\dagger&=&j_a,\label{algjad}\\
{}[j_a,j_b]&=&-i\epsilon_{abc}j_c,\\
{}[j,U]&=&\frac{1}{2}\sigma.U,\label{algj1}\\
{}[j_a, J_b]&=&0,\label{algjJ}\\ 
{}j\cdot j&=&J\cdot J.\label{algj2}
\end{eqnarray}
Direct proofs of these relations, based on definition Eq. \eqref{jfromJU} and
relations \eqref{algJad}, \eqref{algJcom}, \eqref{algJU}, \eqref{algU}, 
can be found in Sect.~\ref{rrotatorapp} of this Supplemental Material. 
Notice that Eqs. \eqref{algjad} and \eqref{algj1} imply that
\(
{}[j,U^*]=-{U^*}.\sigma/2.
\)

\subsection{Bond-algebraic Duality Transformation}

This section describes the construction of a dual representation of 
the Hamiltonian \(H_{\sf PCh}\)
of Eq. \eqref{hpch}. The starting point is the selection of a suitable set 
of bonds as generators
of the bond algebra of interactions. One convenient choice is
\begin{eqnarray}
J_m,\qquad m&=&1,\cdots,N\label{bJ}\\
U_{m+1}^*.U_m,\ U_m^*.U_{m+1}, \qquad m&=&1,\cdots,N-1, \label{buu}
\end{eqnarray}
We call the algebra they generate  \(\mathcal{A}_{\sf PCh}\). Notice that
\(H_{\sf PCh}\in \mathcal{A}_{\sf PCh}\), but the bond algebra
does not include the position-like operators  \(U_m, U_m^*,\ m=1,\cdots,N\).
It will be useful later to change this by adding a boundary term
\begin{equation}\label{bc}
U_N, \qquad U_N^*,
\end{equation}
to the list of generators of \(\mathcal{A}_{\sf PCh}\). The resulting 
extended algebra, still denoted by \(\mathcal{A}_{\sf PCh}\),
does include the \(U_m, U_m^*,\ m=1,\cdots,N\), since
\begin{eqnarray}
U_m&=&U_N.(U_N^*.U_{N-1}).\cdots.(U_{m+1}^*.U_m),\label{Ustring}\\
U_m^*&=&(U_m^*.U_{m+1}).\cdots.(U_{N-1}^*.U_N).U_N^*\ . \label{U*string}
\end{eqnarray}

The extended algebra \(\mathcal{A}_{\sf PCh}\) is simply a direct product
of \(N\) copies of the algebra generated by a single rigid rotator \(J, U, U^*\).
However, what is required is an understanding of the structure of \(\mathcal{A}_{\sf PCh}\)
from the point of view of the local interaction terms in \(H_{\sf PCh}\).
The relations (other than commutation) between the bond generators of Eqs.
\eqref{bJ}, \eqref{buu}, and \eqref{bc} are \(U_N^*.U_N=\mathds{1}\),
\((U_{m+1}^*.U_m).(U_m^*.U_{m+1})=\mathds{1}\), \([J_{m,a},\ J_{n,b}]= 
i\epsilon_{abc} J_{m,c} \, \delta_{m,n}\) for \(m=2,\cdots,N-1\),
\begin{eqnarray}
{}[J_m,U_m^*.U_{m+1}]=-\frac{1}{2}\sigma.U_m^*.U_{m+1},\\
{}[J_m,U_{m-1}^*.U_{m}]=\frac{1}{2}U_{m-1}^*.U_{m}.\sigma\ ,
\end{eqnarray}
and at the boundaries, \(
[J_1,U_1^*.U_{2}]=-\frac{1}{2}\sigma.U_1^*.U_2\),
\([J_N,U_{N-1}^*.U_N]=\frac{1}{2}U_{N-1}^*.U_{N}.\sigma\),
and \([J_N, U_N^*]=-\frac{1}{2}\sigma.U_N^*\).
Relations that follow by Hermitian conjugation from those listed
have been omitted.

\begin{figure}
\centering
\includegraphics[angle=0, width=.9\columnwidth]{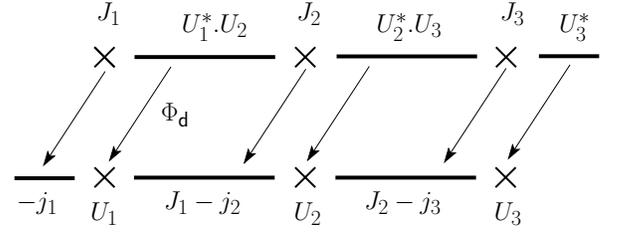}
\caption{Duality automorphism for the quantum chain of rigid rotators,
shown for three sites ($N=3$).}
\label{dsu2I}
\end{figure}

The goal is to construct a mapping that preserves these algebraic relations
and locality. For instance (see Fig.~\ref{dsu2I}),
\begin{eqnarray}
U_N^*&\mapsto& U_N, \qquad U_{m-1}^*.U_m\mapsto U_{m-1}, \label{uudual}\\
J_1&\mapsto& -j_1, \qquad J_m\mapsto -j_m+J_{m-1},  \label{nadmB}
\end{eqnarray}
for \(m=2,\cdots,N.\)
It is not necessary to specify the action of this mapping
on the \(j_m\), since the \(j_m\) are functions of \(J_m,U_m,U_m^*\)
(see Eq. \eqref{jfromJU}).

As noted in the accompanying paper, to verify that the bond-algebra
mapping defined above is induced by a unitary map, we can invoke the
Stone-von Neumann-Mackey theorem~\cite{rosenberg_j1:qc2004a}.  In order
to do so, we need to verify that the operators of the elementary
degrees of freedom are transformed into operators of a covariant pair
of representations as required by the theorem. We can express the
images of the operators in terms of the bonds (including the boundary
terms) directly. A benefit of doing so is that these images define
collective modes of interest.

The dual momenta \(\hat{J}_m\) are by definition
the image \(J_m\mapsto\hat{J}_m\), and are obtained
directly from Eq. \eqref{nadmB},
\begin{eqnarray}
\hat{J}_1=-j_1, \qquad \hat{J}_m=J_{m-1}-j_m,
\end{eqnarray}
for \(m=2,\cdots,N\).
To compute the dual position-like operators it is necessary to exploit
the decompositions of Eqs. \eqref{Ustring} and \eqref{U*string}. These
decompositions combined with Eq. \eqref{uudual} yield
\begin{eqnarray}
U_m\mapsto U_N^*.\cdots.U_m^*\equiv\hat{U}_m,\label{Uhat},
\end{eqnarray}
and \(U_m^*\mapsto U_m.\cdots.U_N\equiv\hat{U}_m^*\).
It can be checked that the dual variables \(\hat{J}_m, \hat{U}_m, \hat{U}_m^*\)
commute on different sites, and satisfy the relations of Eqs. \eqref{algJad}, \eqref{algJcom},
\eqref{algJU}, and \eqref{algU}, as required for a covariant pair of representations.  
 
Similarly to the dual variables, 
the dual Hamiltonian is computed as \(H_{\sf PCh}\mapsto H_{\sf PCh}^D\).
Hence
\begin{eqnarray}
\!\!\! \!\!\! \!\!\! H_{\sf PCh}^D=\frac{1}{2}j_1^2+\!\! \sum_{m=1}^{N-1}\Big[ \frac{1}{2}(j_{m+1}-J_{m})^2 +
\frac{\lambda}{2}{\sf Re}\,\trr(U_{m})\Big]\label{dexp}.
\end{eqnarray}

To gain insight into the physical meaning of Eq. \eqref{dexp} it is useful to
discuss the global symmetries of \(H_{\sf PCh}\) and their
dual representation. On one hand, the interaction terms
${\sf Re}\,\trr({U}_{m+1}^*.{U}_m)$, $m=1,\cdots,N-1$,
are invariant under right and left multiplication, \(U_m\rightarrow{U}_m.v\) and 
\(U_m\rightarrow v.{U}_m\). It follows that \(H_{\sf PCh}\) has a global 
\(SU(2)\times SU(2)\) symmetry, with infinitesimal generators
\(J\equiv\sum_{m=1}^N J_m\) and \( j\equiv\sum_{m=1}^N j_m\) that 
commute with \(H_{\sf PCh}\).
On the other hand, the dual Hamiltonian \(H_{\sf PCh}\) contains 
the terms
${\sf Re}\,\trr(U_m)$, $m=1,\cdots,N-1$, 
invariant only under the adjoint (anti)action, 
\(
\trr(v^*.U_m.v)=\trr(U_m).\label{aa}
\) It may seem that a symmetry has been lost. 

The duality maps the symmetry generators \(j,J\) to dual symmetry generators
\begin{eqnarray}
\hat{J}=\sum_{m=1}^N\hat{J}_m=-j_1+\sum_{m=2}^{N}(-j_m+J_{m-1}),\\
\hat{j}=\sum_{m=1}^N\hat{j}_m=\sum_{m=1}^N\frac{1}{2}
\trr(\hat{U}_m^*.\sigma.\hat{U}_m.(\sigma\cdot\hat{J}_{m})).
\end{eqnarray}
The Hamiltonian \(H_{\sf PCh}^D\) commutes with \(\hat{j},\hat{J}\)
by construction (the duality mapping preserves {\it all} algebraic relations),
meaning that no symmetry has been lost. 
Notice that \(\hat{j}\) presents a highly non-local structure in terms of 
\(J_m, U_m, U_m^*\).

\subsection{Further Results on the Algebra of a Single Quantum Rigid Rotator}
\label{rrotatorapp}

Next it is shown that the operators \(j_a\) defined in Eq. \eqref{jfromJU}
satisfy the relations listed in Eqs. \eqref{algj1} and \eqref{algj2}.
The first step is to introduce the adjoint representation of the \(SU(2)\)
Lie algebra via its double-covering homorphism $R$ to \(SO(3)\),
\(
U\ \mapsto\ R(U),
\)
defined implicitly by
\begin{equation}\label{R}
U.\sigma_a.U^*=\sum_b \sigma_bR(U)^b_{\ \ a}.
\end{equation}
Since \(\trr(\sigma_a . \sigma_b)/2=\delta_{ab}\), \(R(U)\) reads
\begin{equation}\label{explicitR}
R(U)^b_{\ \ a}=\trr(U. \sigma_a . U^* . \sigma_b)/2.
\end{equation}
It follows that Eq. \eqref{jfromJU} can be rewritten as 
\(
j_a= \sum_b R(U^*)^b_{\ \ a}J_b.
\)

{}From Eq. \eqref{explicitR}, 
\begin{eqnarray}\label{ofvec}
[J_a, R(U^*)^b_{\ \ c}]= i\epsilon_{abd}R(U^*)^d_{\ \ c}.
\end{eqnarray}
Also,
\(
[j_a, U]=\frac{1}{2}\sum_b U.\sigma_bR(U^*)^b_{\ \ a}=\frac{1}{2}\sigma_a . U,
\)
where the last equality follows from Eq. \eqref{R} ( 
the conjugate relation \([j_a,U^*]=-U^*. \sigma_a/2\)
follows in the same way). Combining this last result with Eq. \eqref{ofvec}
gives 
\(
[j_b, J_a]=
i\sum_{c,d}(-\epsilon_{adc}+\epsilon_{cad})R(U^*)^c_{\ \ b}J_d=0
\), and
\begin{eqnarray}
&&j\cdot j=\sum_a(\sum_b J_b R(U^*)^b_{\ \ a}) (\sum_c R(U^*)^c_{\ \ a}J_c)\\
&&+\sum_a(\sum_b [R(U^*)^b_{\ \ a},J_b] ) (\sum_c R(U^*)^c_{\ \ a}J_c)
=J\cdot J,\nonumber
\end{eqnarray}
where  the homomorphism property of \(R\) was used to simplify
\(
\sum_a R(U^*)^b_{\ \ a}R(U^*)^c_{\ \ a}=\delta^b_{\ \ c}. 
\)

To check the commutator \([j_a,j_b]\), direct computation gives
\begin{eqnarray}
[j_a,j_b]=\label{lagainstl}-i\epsilon_{dec}R(U^*)^d_{\ \ a}R(U^*)^e_{\ \ b}J_c.
\end{eqnarray}
Since \(R(U)\in SO(3)\), \(\det R(U)=1\). It follows that 
\(
\epsilon_{dec}R(U^*)^d_{\ \ a}R(U^*)^e_{\ \ b}=
\epsilon_{abf}R(U^*)^c_{\ \ f}.
\)
Then Eq. \eqref{lagainstl} simplifies to read
\begin{equation}
[j_a,j_b]=
-i\epsilon_{abf} \sum_c R(U^*)^c_{\ \ f}J_c=-i\epsilon_{abc}\ j_c\ .
\end{equation}

It is only left to show that
\(j_a^\dagger=\frac{1}{2}\sum_{b}\ ([J_b, \trr(U^*. \sigma_a . U . \sigma_b)]+
\trr(U^*. \sigma_a . U . \sigma_b)J_b)= j_a\).
The commutator  vanishes by virtue of Eq. \eqref{ofvec}.

\end{document}